\documentclass[epj]{svjour}
\usepackage{latexsym}
\usepackage{graphics}
\usepackage[latin1]{inputenc}
\usepackage{rotating}
\begin{document}
\renewcommand{\textfraction}{0.00000000001}
\renewcommand{\floatpagefraction}{1.0}
\title{Coherent photoproduction of {\boldmath{$\pi^0$}}- and 
{\boldmath{$\eta$}}-mesons off {\boldmath{$^7$}}Li}
\author{
  Y. Maghrbi\inst{1},
  B.~Krusche\inst{1},   
  J.~Ahrens\inst{2},
  J.R.M.~Annand\inst{3},
  H.J.~Arends\inst{2},
  R.~Beck\inst{2,4},
  V.~Bekrenev\inst{5},
  B.~Boillat\inst{1},
  A.~Braghieri\inst{6},
  D.~Branford\inst{7},
  W.J.~Briscoe\inst{8},
  J.~Brudvik\inst{9},
  S.~Cherepnya\inst{10},
  R.F.B.~Codling\inst{3},
  E.J.~Downie\inst{2,3,8},
  P.~Drexler\inst{11},
  L.V.~Fil'kov\inst{10},
  A.~Fix\inst{12},
  D.I.~Glazier\inst{7},
  R.~Gregor\inst{11},
  E.~Heid\inst{2},
  D.~Hornidge\inst{13},
  I. Jaegle\inst{1},
  O.~Jahn\inst{2},
  V.L.~Kashevarov\inst{10,2},
  I. Keshelashvili\inst{1},
  A.~Knezevic\inst{14},
  R.~Kondratiev\inst{15},
  M.~Korolija\inst{14},
  D.~Krambrich\inst{2}, 
  M.~Lang\inst{2,4},
  V.~Lisin\inst{15},
  K.~Livingston\inst{3},
  S.~Lugert\inst{11},
  I.J.D.~MacGregor\inst{3},
  D.M.~Manley\inst{16},
  M.~Martinez\inst{2},
  J.C.~McGeorge\inst{3},
  D.~Mekterovic\inst{14},
  V.~Metag\inst{11},
  B.M.K.~Nefkens\inst{9},
  A.~Nikolaev\inst{2,4},
  R.~Novotny\inst{11},
  M.~Ostrick\inst{2},
  P.~Pedroni\inst{6},
  F.~Pheron\inst{1},
  A.~Polonski\inst{15},
  S.~Prakhov\inst{9},
  J.W.~Price\inst{9},
  G.~Rosner\inst{3},
  M.~Rost\inst{2},
  T.~Rostomyan\inst{6},
  S. Schadmand\inst{11},
  S.~Schumann\inst{2,4},
  D.~Sober\inst{17},
  A.~Starostin\inst{9},
  I.~Supek\inst{14},
  C.M.~Tarbert\inst{7},
  A.~Thomas\inst{2},
  M.~Unverzagt\inst{2,4},
  D.P.~Watts\inst{7},
  D.~Werthm\"uller\inst{1},
  F.~Zehr\inst{1}
\newline(The Crystal Ball at MAMI, TAPS, and A2 Collaborations)
\mail{B. Krusche, Klingelbergstrasse 82, CH-4056 Basel, Switzerland,
\email{Bernd.Krusche@unibas.ch}}
}
\institute{Department of Physics, University of Basel, Ch-4056 Basel, Switzerland
  \and Institut f\"ur Kernphysik, University of Mainz, D-55099 Mainz, Germany
  \and School of Physics and Astronomy, University of Glasgow, G12 8QQ, United Kingdom
  \and Helmholtz-Institut f\"ur Strahlen- und Kernphysik, University of Bonn, D-53115 Bonn, Germany
  \and Petersburg Nuclear Physics Institute, RU-188300 Gatchina, Russia
  \and INFN Sezione di Pavia, I-27100 Pavia, Pavia, Italy
  \and School of Physics, University of Edinburgh, Edinburgh EH9 3JZ, United Kingdom
  \and Center for Nuclear Studies, The George Washington University, Washington, DC 20052, USA
  \and University of California Los Angeles, Los Angeles, California 90095-1547, USA
  \and Lebedev Physical Institute, RU-119991 Moscow, Russia
  \and II. Physikalisches Institut, University of Giessen, D-35392 Giessen, Germany
  \and Laboratory of Mathematical Physics, Tomsk Polytechnic University, Tomsk, Russia
  \and Mount Allison University, Sackville, New Brunswick E4L3B5, Canada
  \and Rudjer Boskovic Institute, HR-10000 Zagreb, Croatia
  \and Institute for Nuclear Research, RU-125047 Moscow, Russia
  \and Kent State University, Kent, Ohio 44242, USA
  \and The Catholic University of America, Washington, DC 20064, USA
}
\authorrunning{Y. Maghrbi et al.}
\titlerunning{Coherent photoproduction of mesons off $^7$Li}

\abstract{Coherent photoproduction of $\pi^0$-mesons from threshold 
($E_{th} \approx$ 136~MeV) throughout the $\Delta$-resonance region 
and of $\eta$-mesons close to the production threshold 
($E_{th} \approx$ 570~MeV for $\eta$) has been measured for $^7$Li nuclei. 
The experiment was performed using the tagged-photon beam of the Mainz MAMI 
accelerator with the Crystal Ball and TAPS detectors combined to give an 
almost 4$\pi$ solid-angle electromagnetic calorimeter. The reactions were 
identified by a combined invariant-mass and missing-energy analysis. A comparison 
of the pion data to plane-wave impulse modelling tests the nuclear mass form 
factor. So far coherent $\eta$-production had been only identified for the 
lightest nuclear systems ($^2$H and $^3$He). For $^3$He a large enhancement 
of the cross section above plane-wave approximations had been reported, 
indicating the formation of a quasi-bound state. The present Li-data for 
$\eta$-production agree with a plane-wave approximation. Contrary to $^3$He, 
neither a threshold enhancement of the total cross section nor a deviation 
of the angular distributions from the expected form-factor dependence 
were observed.
\PACS{
      {13.60.Le}{Meson production}   \and
      {14.20.Gk}{Baryon resonances with S=0} \and
      {25.20.Lj}{Photoproduction reactions}
            } 
} 
\maketitle

\section{Introduction}

Photoproduction of mesons off nuclei involves in general many different final 
states of the meson-nucleus system and can contribute to a wide range of topics
(see \cite{Krusche_11} for an overview). Very interesting for many questions
are two limiting cases. In `quasi-free' processes, the reaction involves one 
specific nucleon, called `participant', which is kicked out of the nucleus, 
and the rest of the nucleus can be regarded as a `spectator' system that only 
compensates the momentum of the bound participant. For light nuclei,
this process is a powerful tool for the study of reactions off quasi-free
neutrons \cite{Krusche_11}; for heavy nuclei it can be used as a testing
ground for meson - nucleus interactions and hadron in-medium properties
\cite{Krusche_04}. 

In `coherent' reactions, ideally the meson is produced via a superposition of 
the reaction amplitudes from all nucleons and, in the final state, the nucleus
remains in its ground state. A similar process in which no nucleon is removed 
from the nucleus but the nucleus is excited to a higher lying nuclear state, is
sometimes called `incoherent' production.
The advantage of the coherent process is the simplicity of the 
final state; the ground-state properties of nuclei are well under control. 
This reaction is well suited for the study of the in-medium properties of
mesons and nucleon resonances. The undisturbed final state can be easily
constructed from the plane-wave impulse approximation and any deviations may be 
attributed to nuclear effects like meson-nucleus final-state interactions or
in-medium modifications of hadron properties. Such programs have been pursued 
in particular for the study of medium effects on the production and propagation 
of the $\Delta$-resonance in medium via the coherent $\gamma A\rightarrow A\pi^0$
reaction (see e.g. \cite{Drechsel_99,Krusche_02}). The same reaction was also
exploited for the study of nuclear properties such as nuclear mass form factors
\cite{Krusche_05}, and in incoherent production, for nuclear transition form
factors \cite{Tabert_08}. Nuclear form factors in the region of helium and
lithium isotopes have gained much new interest in connection with the study
of halo nuclei (see e.g. \cite{Tomaselli_00,Sick_11}). Coherent pion photoproduction 
allows the direct study of the nuclear mass distribution because production of 
$\pi^0$ mesons in the $\Delta$-resonance region couples identically to protons 
and neutrons. 

Until now, coherent photoproduction of heavier mesons off nuclei has almost not been 
investigated since such measurements are very demanding. For mesons like the 
$\eta$, large momenta are transfered to the nucleus, which suppresses the production 
cross section due to the nuclear form factors. Background from breakup reactions, 
where the participating nucleon is removed from the nucleus, dominates the 
production process. This background must be suppressed either by detection of the 
recoil nucleons or by conditions on the reaction kinematics, demanding detector 
systems with large solid-angle coverage, large detection efficiency, and excellent 
energy and angular resolution. 

Recently, coherent photoproduction of $\eta$-mesons from light nuclei 
has attracted interest as a tool for the search of so-called $\eta$-mesic nuclei
\cite{Pfeiffer_04,Pheron_12}.
The question is whether the strong interaction allows the formation of quasi-bound
meson-nucleus states, which would be the ideal system for the study of meson-nucleus
interactions. The interaction of low-energy pions with nuclei is too weak for
quasi-bound states but the situation is much different for $\eta$-mesons.          
Production of $\eta$-mesons in the threshold region is dominated by the excitation
of the s-wave S$_{11}$(1535) resonance \cite{Krusche_95,Krusche_97}, which couples 
strongly (branching ratio $\approx$ 50\% \cite{PDG}) to $N\eta$. As a consequence,
the interaction of $\eta$-mesons with nuclear matter is important also for very 
small momenta of the mesons. Typical absorption cross sections are around 30 mb and 
are over a wide range of kinetic energy ($T\approx$ 1~MeV - 1~GeV) almost independent 
of $T$ \cite{Roebig_96,Mertens_08}. First evidence for an attractive s-wave $\eta N$
interaction, which might lead to the formation of quasi-bound states,
was reported from coupled channel analyses of pion-induced $\eta$-production
reactions \cite{Bhalerao_85,Liu_86} in the 1980s. However, it is still 
controversially discussed whether the interaction is strong enough to form such
states. The original prediction was for nuclei with mass numbers $A$ in the range
slightly above 10. However, refined values for the $\eta N$-scattering length 
extracted from more precise recent $\eta$-production data extended the discussion
to very light nuclei like hydrogen and helium isotopes. (See \cite{Pheron_12} and 
refs. therein for a summary of recent results.)   

A much explored experimental approach to identify $\eta$-mesic states is the study
of the threshold behavior of $\eta$-production reactions. Quasi-bound states in the
vicinity of the production threshold should give rise to an enhancement of the 
respective cross section over phase-space behavior. Many hadron-induced reactions
(see refs. in \cite{Pheron_12}) have been studied for this purpose. Interesting
threshold effects have been observed for many of them. Particularly strong
enhancements were found for the $pd\rightarrow\eta  ^3\mbox{He}$ \cite{Mayer_96} and
$dp\rightarrow\eta  ^3\mbox{He}$ reactions \cite{Smyrski_07,Mersmann_07,Rausmann_09}, 
implying a large $\eta^3$He scattering length. If such effects
are due to a resonance in the $\eta$-nucleus system, they should exist independently
on the initial state of the reaction. 

Electromagnetic induced reactions, like photoproduction of mesons, offer a very 
clean way to study the $\eta$-nucleus final state, but have small production cross 
sections, in particular for the coherent process. Photoproduction of $\eta$-mesons 
in the threshold region has been studied for several hydrogen and helium isotopes 
\cite{Pfeiffer_04,Krusche_95,Krusche_95a,Hoffmann_97,Weiss_01,Weiss_03,Hejny_99,Hejny_02}
and these results allowed the characterization of the spin and isospin
structure of the relevant transition amplitudes \cite{Krusche_03}. The reaction is 
dominated by the excitation of the S$_{11}$(1535) resonance via the 
$E_{0+}$-multipole, which involves a spin-flip of the participating nucleon. 
This means, that coherent $\eta$-production is practically forbidden for nuclei 
with spin $J=0$ ground states. Also for nuclei with non-zero ground-state spins,
depending on the nuclear structure, only a fraction of the nucleons (those which 
can participate in spin-flip transitions) may contribute. Furthermore, the 
electromagnetic excitation of the S$_{11}$(1535) resonance is mainly isovector 
($A_{1/2}^{IS}/A_{1/2}^p \approx 0.1$, where $A_{1/2}^p$ is the helicity coupling for
the proton and $A^{IS}_{1/2}$ is the isoscalar part of the helicity coupling) 
\cite{Krusche_03}, so that contributions from protons and neutrons will cancel to 
a large extent in coherent $\eta$-production. Together with the large momentum 
transfers involved, these features lead to very small reaction cross sections.

\begin{figure}[thb]
\centerline{
\resizebox{0.45\textwidth}{!}{%
  \includegraphics{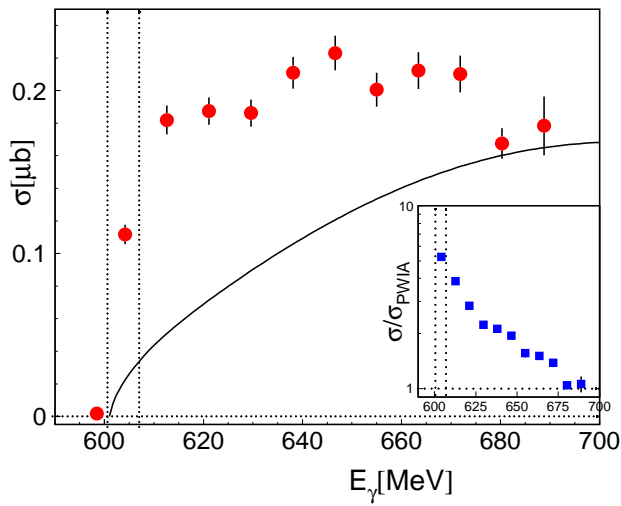}
}}
\caption{Total cross section for the $\gamma ^3\mbox{He}\rightarrow \eta^3\mbox{He}$
reaction \cite{Pheron_12} compared to plane-wave impulse approximation. 
Vertical dashed lines indicate coherent and breakup threshold for $\eta$-production.
Insert: ratio of data and impulse approximation.
}
\label{fig:helium}       
\end{figure}

Only nuclei with ground-state spin $J$ and isospin $I$ different from zero are 
promising candidates for the observation of the coherent process. Previous 
experimental results are consistent with this picture. The cross section for 
coherent production off the deuteron ($J$=1, $I$=0) is small \cite{Weiss_01}, 
(typical values for $d\sigma/d\Omega$ are on the order of 10 nb/sr). Only upper limits 
have been extracted for the $J=I=0$ nucleus $^4$He \cite{Hejny_99}. The most 
interesting case studied so far is the $J=I=$1/2 nucleus $^3$He 
\cite{Pfeiffer_04,Pheron_12}. The coherent process was clearly identified.
The energy dependence of the total cross section shown in Fig.~\ref{fig:helium}
\cite{Pheron_12} is different from the expectation for reaction phase-space. 
A strong threshold enhancement relative to the plane-wave impulse approximation (PWIA), 
similar to the results from hadron-induced reactions        
\cite{Smyrski_07,Mersmann_07,Rausmann_09}, is observed. The angular distributions 
close to threshold are more isotropic than expected from the shape of the 
nuclear form factor \cite{Pheron_12}. Both observations together have been taken 
as indication for the formation of a resonant-like meson-nucleus state 
\cite{Pfeiffer_04,Pheron_12}.

So far, this is the only isolated case where coherent $\eta$-threshold production 
off nuclei could be studied. Almost nothing is known experimentally about its 
systematics and the validity of the simple plane-wave impulse approximation used 
in \cite{Pfeiffer_04,Pheron_12}.
The present work therefore aimed at the measurement of this reaction from a different 
light nucleus. Apart from $^3$H, the mirror nucleus of $^3$He, which, however, 
is difficult to handle as a target, the lightest stable isotope with nonzero 
ground-state spin ($J^{\pi}=$ 3/2$^-$) and isospin ($I=$ 1/2) is $^7$Li. In the 
relevant range of momentum transfer its squared form factor \cite{Suelzle_67}, 
which is expected to be proportional to the cross section, is roughly smaller by 
an order of magnitude compared to $^3$He \cite{McCarthy_77}. However, a factor of 
$\approx$ 3 in counting statistics may be recovered from the target thickness 
(number of nuclei/cm$^2$), making the measurement feasible.

This paper is organized as follows. The assumptions and inputs for the 
modelling of coherent $\pi^0$- and $\eta$-pho\-to\-pro\-duc\-tion off $^7$Li in
plane-wave impulse approximation are discussed in Sec.~\ref{sec:pwia}. 
The experimental setup is described in Sec. \ref{sec:setup}
and the data analysis, in particular the identification of events from the coherent
process, is discussed in Sec. \ref{sec:ana}. The measured cross sections for coherent 
$\pi^0$ and $\eta$-production are summarized in Sec. \ref{sec:results} and compared 
to the results of the PWIA modelling. 
    
\section{Plane wave impulse approximation}
\label{sec:pwia}

\begin{figure}[htb]
\resizebox{0.49\textwidth}{!}{%
  \includegraphics{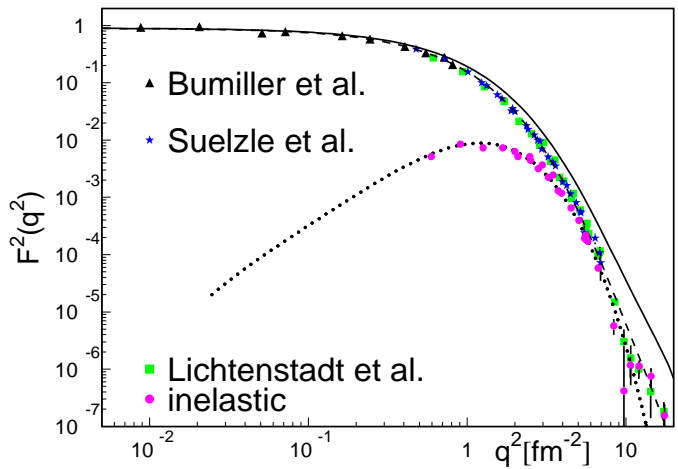}
}
\caption{Elastic charge form factors from Bumiller et al. \cite{Bumiller_72}, 
(black) triangles, Suelzle et al.\cite{Suelzle_67}, (blue) stars, 
and Lichtenstadt et al. (green) squares.
Dashed line: parametrization of form factor $F_{C}$, Solid line: form factor 
$F_{C*}$.  (Magenta) dots: inelastic form factor for 478-keV excitation
\cite{Lichtenstadt_89}. Dotted curve: parametrization of inelastic form factor. 
}
\label{fig:chff}       
\end{figure}

The PWIA of the coherent meson production follows the work of Drechsel et al. 
\cite{Drechsel_99}, taking into account the specific features of the $\pi^0-A$
and $\eta-A$ final states. The main inputs are nuclear form factors and
the amplitudes for the elementary meson production reactions off the free nucleon.
The elastic charge form factor $F_{C}$ of $^7$Li has been measured with electron 
scattering over a wide range of momentum transfer $q$ 
\cite{Suelzle_67,Bumiller_72,Lichtenstadt_89}. 
Lichtenstadt et al. \cite{Lichtenstadt_89} also reported results for the inelastic 
transition form factor $F_{Cx}$ related to the excitation of the 478-keV state in 
$^7$Li. Since the charge and mass rms radii of $^7$Li are similar 
\cite{Tomaselli_00,Kajino_88}, we can use the charge form factors as basis. However, 
they include the effects from the charge distribution of the proton. For the meson 
production reactions we need instead the distribution of point-like nucleons.
Therefore, the measured charge form factors must be divided by the proton dipole 
form factor $F_p^2(q^2)$; the ratios are denoted by $F_{C*}$ and $F_{Cx*}$. 
Figure \ref{fig:chff} summarizes the charge form factors and their parametrizations
used in the PWIA modelling. For the elastic form factor, the parametrization
of $F_{C*}$ is also shown. The $q$-dependence of the inelastic form factor $F_{Cx}$ 
for small values of $q$ is approximated by the model results cited in 
\cite{Lichtenstadt_89}.

The construction of the transition amplitudes starts from the effective total 
energy $W=\sqrt{s^{\rm eff}}$ of the incident photon (four-momentum $P_{\gamma}$, 
laboratory energy $E_{\gamma}$) and an off-shell nucleon (four-momentum $P_{N}$) 
with three-momentum $\vec{p}_{N}$ from its motion inside the nucleus
\begin{equation}
s_{\rm eff} = (P_{\gamma} + P_{N})^2.
\end{equation}
The nucleon momentum $\vec{p}_N$ is obtained in the factorization approximation 
\cite{Drechsel_99} from the momentum transfer $\vec{q}$ to
the nucleus by
\begin{equation}
{\bf{p}}_N = -\frac{A-1}{2A}{\bf{q}} = -\frac{3}{7}{\bf{q}}\; ,
\end{equation}
where $A$ is the nuclear mass number and all momenta are in the laboratory frame
(note that the expressions in \cite{Drechsel_99} refer to the center-of-momentum frame). 
The amplitudes of the elementary reactions are then evaluated at 
$W(E_{\gamma},\vec{q})$. 

The amplitude for meson photoproduction off nuclei is in general given by
\begin{equation}
{\cal{F}} = L +i \sigma K, \;\;\; d\sigma = |L|^2 +|K|^2 
\end{equation}
with the spin-independent part $L$ and the spin-dependent part $K$. It is 
efficiently evaluated in the CGLN parameterization \cite{Chew_57}, involving the 
four invariant amplitudes $F_1$,...,$F_4$. 

The simplest case is coherent $\pi^0$-production from spin $J=0$ nuclei in the 
$\Delta$-resonance region \cite{Drechsel_99}. The elementary production amplitudes 
are identical for protons and neutrons. The dominant contribution to 
$\gamma A\rightarrow \pi^0 A$ for spin $J=0$ nuclei thus involves the 
spin/isospin-independent part of the production amplitude. 
In the CGLN representation a spin-independent piece arises only from the term
with the $F_2$ amplitude. Due to the pseudoscalar nature of the pion and the 
overall symmetry of the problem, this term has a $\mbox{sin}(\Theta^{\star}_{\pi})$ 
factor ($\Theta_{\pi}^{\star}$: pion polar angle in photon-nucleus cm-system) in 
the amplitude \cite{Drechsel_99}. Since the dominant excitation of the 
$\Delta$-resonance is not isospin dependent, all amplitudes from protons and 
neutrons add coherently, which is reflected in a factor $A$ in the amplitude.
The full evaluation of the $L$-piece gives:
\begin{equation}
\frac{d\sigma_0}{d\Omega} = \frac{1}{2} \frac{q_{\pi}^{\star}}{k_{\gamma}^{\star}}
  |F_{2}(W)|^2 A^2 \mbox{sin}^2(\Theta_{\pi}^{\star}) F^2_{C*}(q^2)\;,
\label{eq:pi_no}
\end{equation} 
where the ratio of pion and photon momenta $q_{\pi}^{\star}$, $k_{\gamma}^{\star}$ 
gives the phase-space factor for the photon-nucleus system. Numerical values for 
the CGLN amplitude $F_{2}$ were taken from the MAID analysis of pion photoproduction
\cite{MAID}. 

The $^7$Li case is complicated by the unpaired proton in the $1p_{3/2}$ orbit,
which gives rise to additional contributions involving also spin-flip amplitudes
that may contribute to all four CGLN amplitudes. Apart from elastic reactions, the 
$1p_{3/2}$ proton may be excited to the $1p_{1/2}$ orbit, populating the low lying 
$1/2^-$ state of $^7$Li with an excitation energy of 478 keV. Incoherent pion 
production to this final state cannot be separated experimentally from the coherent 
process and is thus included in the measured cross sections. The spin-dependent 
contribution must be small compared to the spin-independent contribution because 
it is lacking the $A^2$ factor, but it is important for extreme forward or backward 
angles (because it has a $\mbox{cos}^2(\Theta_{\pi}^{\star})$ dependence instead of the 
$\mbox{sin}^2(\Theta_{\pi}^{\star})$ for the spin-independent part). 
These contributions are approximated from the leading $M_{1+}$ multipole. 
Evaluation of the multipole expansion of the CGLN amplitudes for the spin-dependent 
part of the cross section leads to 
\begin{equation}
\frac{d\sigma_{sf}}{d\Omega} \approx \frac{q_{\pi}^{\star}}{k_{\gamma}^{\star}}
 |M_{1+}(W)|^2 \mbox{cos}^2(\Theta_{\pi}^{\star}) \left(F^2_{C*}(q^2) +F^2_{Cx*}(q^2)\right)
\label{eq:pi_sf}
\end{equation}
when all multipoles except the leading $M_{1+}$ are neglected. The incoherent 
excitation of the nucleus is included, but the contribution turns out to be 
negligible (see Sec.~\ref{sec:pires}). The amplitudes $M_{1+}$ are again taken from 
the MAID-model \cite{MAID}. For the full PWIA cross section the incoherent sum
\begin{equation}
\frac{d\sigma_{\pi A}}{d\Omega} = \frac{d\sigma_{0}}{d\Omega} 
+ \frac{d\sigma_{sf}}{d\Omega} 
\label{eq:pi_pwia}
\end{equation}
is used.

The situation is different for $\eta$-production. Since the elementary reaction is 
completely dominated by an isovector, spin-flip amplitude, there is no piece 
corresponding to Eq.~\ref{eq:pi_no} in pion production. Like in the $^3$He case 
\cite{Pheron_12}, the main contribution to coherent production comes from the 
S$_{11}$ excitation of the unpaired nucleon via a spin-flip transition. The main 
difference is that $^3$He has an unpaired neutron while $^7$Li has an unpaired 
proton. We use therefore a similar PWIA as for $^3$He in 
\cite{Pheron_12}, including the incoherent excitation via the $F_{Cx*}$-term
\begin{equation}
\frac{d\sigma_{\eta A}}{d\Omega} = 
\left(\frac{q_{\eta}^{(A)}}{k_{\gamma}^{(A)}}
\frac{k_{\gamma}^{(N)}}{q_{\eta}^{(N)}}\right) 
\frac{d\sigma_{\rm elem}}{d\Omega}
\left(F^2_{C*}(q^2) + F^2_{Cx*}(q^2)\right)
\label{eq:eta}
\end{equation}   
with a parameterization of the measured $\gamma p\rightarrow p\eta$ cross section
from \cite{McNicoll_10} for the elementary cross section $d\sigma_{\rm elem}$.
The change of phase space between the different c.m. systems is derived from the 
photon and $\eta$ three-momenta in the photon-nucleon 
($k_{\gamma}^{(N)}$, $q_{\eta}^{(N)}$), 
and photon-nucleus ($k_{\gamma}^{(A)}$, $q_{\eta}^{(A)}$) c.m. systems.

\section{Experimental setup}
\label{sec:setup}
The experimental setup was identical to the one used in \cite{Schumann_10,Zehr_12}, apart
from the target (liquid hydrogen targets for \cite{Schumann_10,Zehr_12}, threshold settings,
and trigger conditions.

\begin{figure}[htb]
\centerline{\resizebox{0.48\textwidth}{!}{%
  \includegraphics{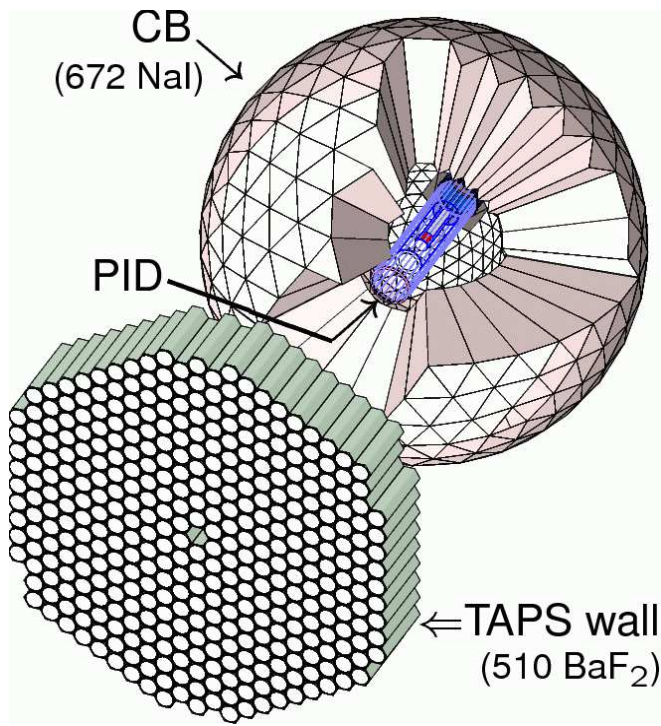}
}}
\caption{Setup of the electromagnetic calorimeter combining the
Crystal Ball and TAPS detectors. Detectors for charged particle identification
were mounted in the Crystal Ball (PID and MWPC) and in front of the TAPS forward 
wall (TAPS Veto-detector).
}
\label{fig:calo}       
\end{figure}

The measurement used the tagged photon beam \cite{Anthony_91,Hall_96} from a primary 
883 MeV electron beam of the Mainz MAMI accelerator \cite{Herminghaus_83,Walcher_90}.
The photons irradiated a $^7$Li target (enrichment 99\%) of 5.4 cm length and a 
density of 0.534 g/cm$^{3}$, corresponding to a surface density of 0.264 
nuclei/barn. The reaction products were detected with an electromagnetic 
calorimeter composed of the Crystal Ball (CB) \cite{Starostin_01} and TAPS detectors
\cite{Novotny_91,Gabler_94}. The 672 NaI crystals of the CB covered the full 
azimuthal angle for polar angles between 20$^{\circ}$ and 160$^{\circ}$ around the 
target, which was mounted in the center of the CB. TAPS covered polar angles between
1$^{\circ}$ and 20$^{\circ}$ as a hexagonal wall of 510 BaF$_{2}$ crystals, mounted 
1.75 m downstream from the target. Individual plastic detectors in front of each 
crystal were used for charged particle identification. A schematic view of the 
setup, which covered $\approx$ 98\% of 4$\pi$, is shown in Fig.~\ref{fig:calo}.
It was complemented by a cylindrical Particle Identification Detector (PID) 
\cite{Watts_04}, mounted around the target inside the CB, which covered the same 
solid angle as the CB.

The experiment trigger was based on a subdivision of the CB and TAPS into logical 
sectors. For TAPS these were eight sectors of 64 modules in a pizza-like geometry, and 
for the CB 45 rectangles. The trigger required signals in at least two logical sectors
of the calorimeter above a threshold of 20~MeV and an analog energy sum of the CB 
modules above 50 MeV. Once a valid trigger had been generated, thresholds for the 
readout of individual modules were 5~MeV in TAPS and 2~MeV in the CB.  

\section{Data analysis}
\label{sec:ana}
The different analysis steps for the identification of photons, charged pions, and 
recoil nucleons are discussed in more detail in \cite{Schumann_10,Zehr_12}. 
The analysis of coherent neutral meson production off nuclei is special in so far
as no charged particles (no charged pions, no recoil protons) may occur in the 
final state. Detection of charged particles was only used to veto events, which 
simplifies the analysis (there was no need to separate charged pions from protons 
or to extract energy information for the charged particles). Accepted were only 
events with exactly two photons (from the $\pi^0\rightarrow \gamma\gamma$ or 
$\eta\rightarrow \gamma\gamma$ decays) or with exactly six photons 
($\eta\rightarrow 3\pi^0\rightarrow 6\gamma$). These are particularly clean
data samples.

\begin{figure}[thb]
\centerline{\resizebox{0.45\textwidth}{!}{%
  \includegraphics{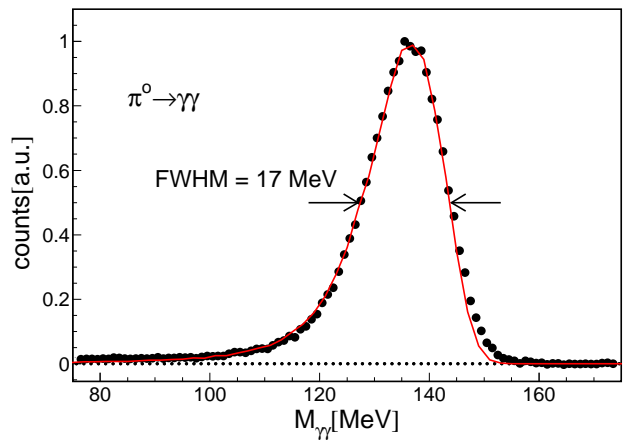}
}}
\caption{Invariant mass spectrum for two-photon events for incident photon energies
below 300 MeV. Statistical uncertainties smaller than symbol sizes.
The solid (red) curve is a Monte Carlo simulation of the detector 
response. 
}
\label{fig:pi_inv}       
\end{figure}

The invariant-mass spectrum of photon pairs for incident photon energies below 
300~MeV is shown in Fig.~\ref{fig:pi_inv}. It is practically background free. 
No other reactions with significant cross section produce two or more photons 
in this energy range. 

\begin{figure}[htb]
\centerline{\resizebox{0.45\textwidth}{!}{%
  \includegraphics{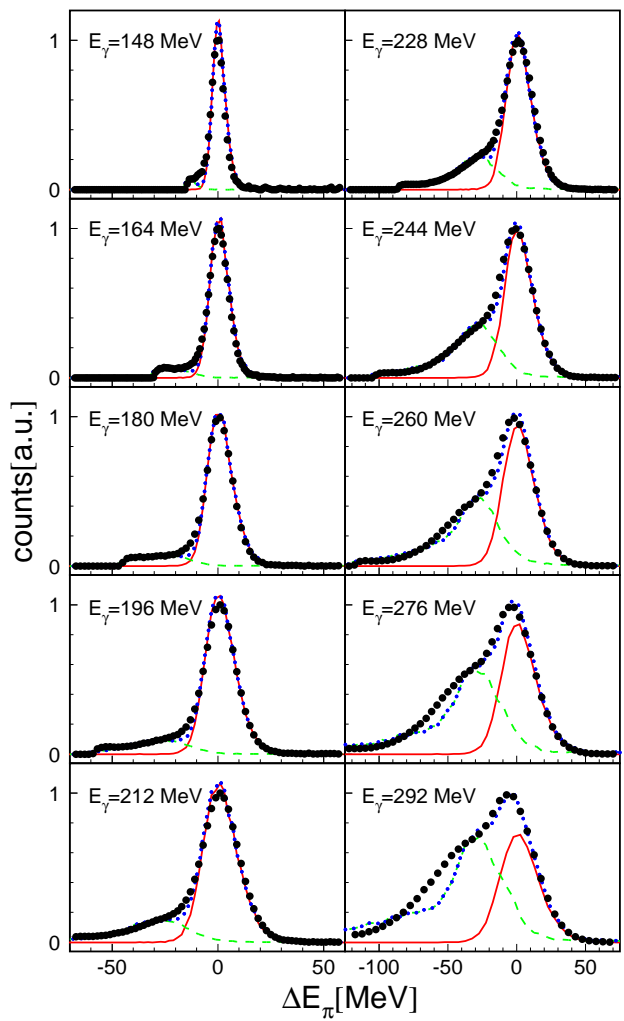}
}}
\caption{Missing energy analysis for single $\pi^0$ production for different  
incident photon energies. Black dots: measurement (statistical uncertainties smaller
than symbol size), 
solid (red) curves: MC-simulation for coherent events, 
dashed (green) curves: MC for breakup events,
dotted (blue): sum of both. 
}
\label{fig:pi_misse}       
\end{figure}

\clearpage

\begin{figure}[thb]
\resizebox{0.43\textwidth}{!}{%
  \includegraphics{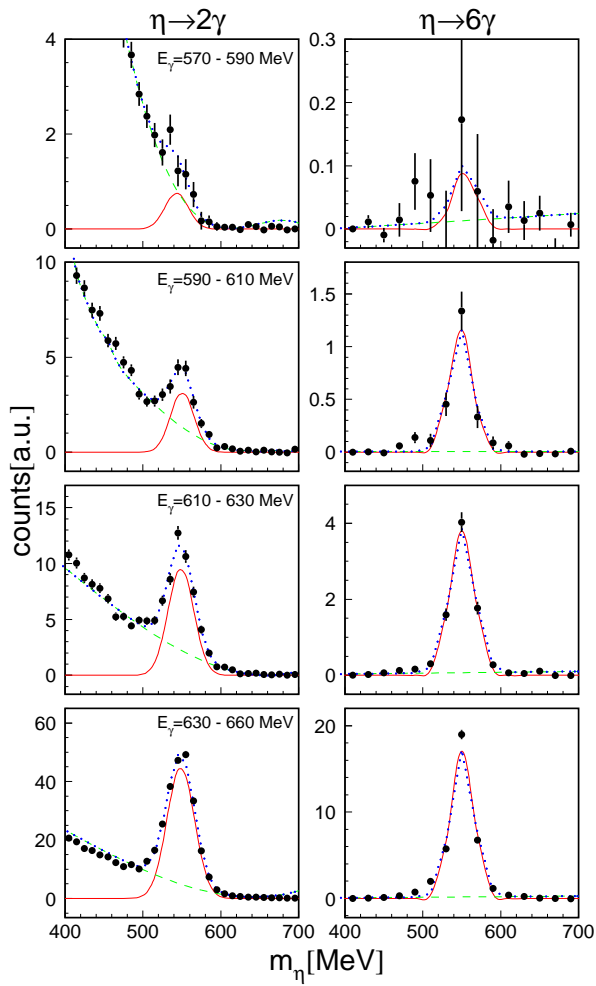}
}
\caption{Invariant-mass spectra for two-photon (left-hand side) and six-photon 
(right-hand side) events in the energy region of the $\eta$-production threshold. 
Solid (red) lines: signal shapes, dashed (green) lines: fitted background, 
dotted (blue) curves: sum of both. Ranges of incident photon energies are given 
at left-hand side.
}
\label{fig:eta_inv}       
\end{figure}

Double $\pi^0$ production sets in with a very low cross section around 300 MeV
and loss of two of the four decay photons is unlikely. The only possible background 
source is production of single $\pi^0$ off quasi-free neutrons with loss of one 
decay photon and misidentification of the neutron as photon. However, the 
corresponding recoil neutrons are mostly emitted to forward angles and can be 
identified in TAPS with time-of-flight versus energy and pulse-shape analyses.  
The important step is then the separation of the coherent reaction from breakup 
reactions with emission of recoil nucleons. The suppression of such events by the 
required non-detection of recoil nucleons is limited since the detection 
efficiency for recoil neutrons is only on the order of 30\% (larger than 90\% 
for recoil protons). We use therefore in addition the overdetermination
of the reaction kinematics of the two-body final state. 
The laboratory kinetic energy of the meson $E_m^{\rm lab}$ is directly measured with 
the calorimeter, and its kinetic cm-energy $E_m^{\star}$ follows from the incident 
photon energy $E_{\gamma}$.

\begin{figure}[thb]
\resizebox{0.45\textwidth}{!}{%
  \includegraphics{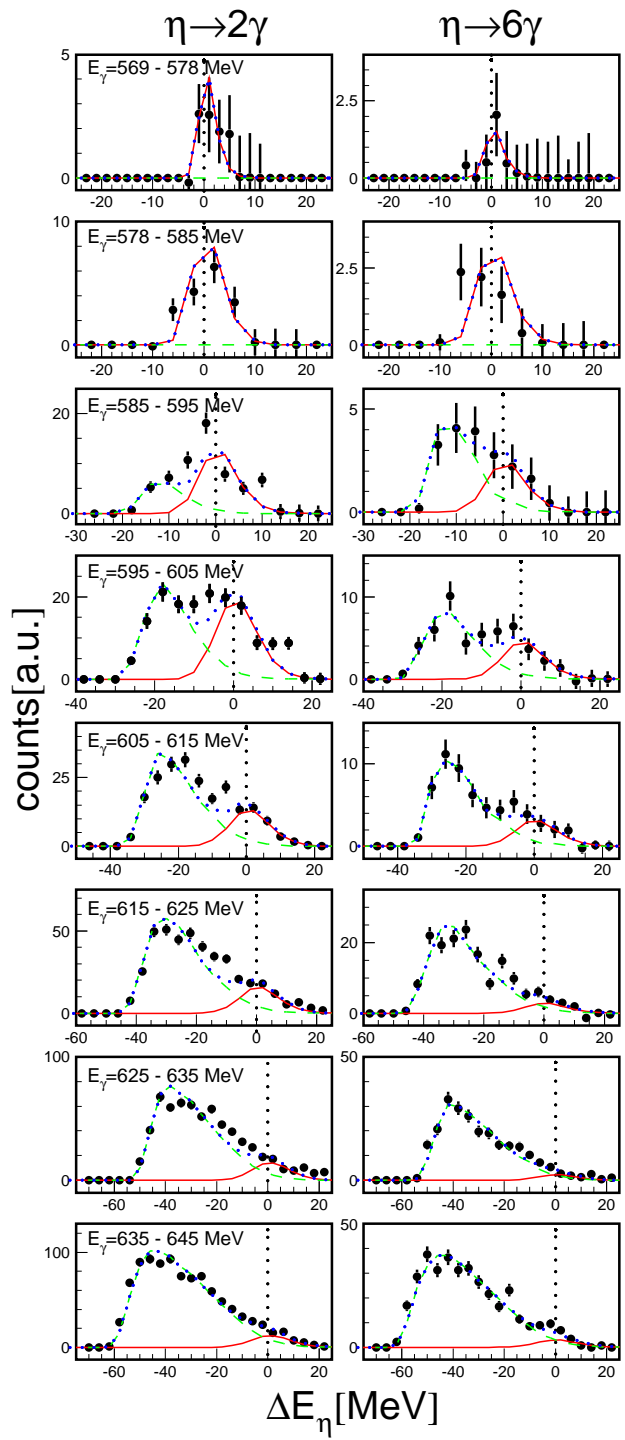}
}
\caption{Missing-energy spectra for events in the $\eta$ invariant mass peaks for 
different ranges of incident photon energy. Notation for curves is as in 
Fig.~\ref{fig:pi_inv}. Vertical dotted lines: expected positions of coherent peaks. 
Left-hand side: two-photon events, right-hand side: six-photon events. 
}
\label{fig:eta_misse}       
\end{figure}

The mesons are boosted into the cm-system and the 
difference $\Delta E_{\pi}$ of the two kinetic energies in the cm-system is 
constructed as
\begin{equation}
\Delta E_m = E_m^{\star}(E_m^{\rm{lab}}) - E_m^{\star}(E_{\gamma}).
\end{equation}  

\clearpage

The result of this analysis for pion production is shown in Fig.~\ref{fig:pi_misse}. 
The peaks at zero missing energy correspond to coherent production and dominate the 
process at low incident photon energies. At higher incident photon energies breakup 
background appears at negative missing energies. The shape of the signals was 
generated with a full Monte Carlo simulation of the experiment using the GEANT3 
package \cite{Brun_86}. The event generator for the coherent process was based on 
trivial two-body kinematics; for the breakup reaction, the momentum distribution of 
the bound nucleons was taken into account. Final state interactions were not taken 
into account, which explains the deviations between data and Monte Carlo in the
tails of the distributions for higher incident photon energies. 
The separation of coherent and breakup 
processes, which must be done in dependence on the pion angles, is straightforward 
for the energy range up to $E_{\gamma}$ = 300 MeV as shown in Fig.~\ref{fig:pi_misse}. 
At higher incident photon energies, the contribution from breakup reactions becomes 
dominant in the angle-integrated missing-energy spectra. Across the angular 
distribution missing energy spectra vary. The fraction of coherent events compared to
breakup is larger for forward angles, but due to kinematics the separation between 
coherent and breakup events in missing energy is better at backward angles.
The `coherent reaction' includes incoherent excitation of the 478-keV level, which cannot
be resolved by the missing energy analysis.

The analysis for coherent $\eta$-production follows the same scheme. Invariant-mass 
and missing-energy spectra are summarized in 
Figs.~\ref{fig:eta_inv} and \ref{fig:eta_misse}. The main difference to $\pi^0$ 
production is that, near threshold, the ratio of coherent to breakup 
cross sections is much less favorable. This comes from two effects discussed in 
Sec.~\ref{sec:pwia}. The involved momentum transfers are much larger,
suppressing the coherent cross section via the form factor. Since furthermore 
(apart from small components in the nuclear wave functions) only the $1p_{3/2}$ 
proton contributes, the $A^2$ factor is missing in the coherent cross section.

The invariant-mass peaks from the two-photon decays show some background 
(double $\pi^0$ production with two undetected photons, single $\pi^0$ production 
off quasi-free neutrons with one undetected photon and a misidentified neutron), 
which must be subtracted. The invariant-mass signals of the six-photon decays are
much cleaner. In this case, the invariant masses of the three $\pi^0$-mesons 
are also used to identify the reaction as discussed in \cite{Pheron_12}. 
The contribution of breakup background to the missing-energy spectra is substantial.
A clean coherent signal appears only in the immediate threshold region. At higher 
energies the signal can be extracted only by fitting the simulated line-shapes to 
the data, which for incident photon energies above 650 MeV becomes unfeasible. 
However, the simultaneous extraction of the cross section from the two different 
$\eta$-decay channels gives some estimate for the typical level of uncertainty.

Absolute cross sections were extracted from the measured yields with the target 
surface density, the incident photon flux, and the simulated detection efficiencies. 
The latter were generated with GEANT3 \cite{Brun_86} simulations. Typical values 
(depending on incident photon energy and polar angle of the meson) are 20\% - 50\% 
for coherent $\pi^0$ production and 35\% - 40\% for coherent $\eta$-production 
to the six-photon final state and 60\% - 70\% for the two-photon final state. 
The uncertainty for the detection efficiency simulations is smaller than in 
\cite{Zehr_12} for two reasons. Only photons had to be detected, for which the 
response of the detector system is best understood. There is no additional uncertainty 
from the properties of the event generator because in both cases only trivial two-body 
kinematics is involved in the final state. We estimate the systematic uncertainty of 
the detection efficiency below the 5\% level. The incident photon flux was determined 
from the counting of the number of deflected electrons in the focal plane by live-time 
gated scalers. The fraction of correlated photons that pass the collimator and reach 
the target (tagging efficiency, $\approx$ 50\% for this experiment) was determined with 
special experimental runs. A total absorbing lead-glass counter was moved into the 
photon beam at reduced intensity of the primary electron beam. The intensity was reduced 
at the electron source, so that no accelerator parameters differed from normal running.
In addition to these periodical absolute measurements the intensity was monitored in 
arbitrary units during normal data taking with an inonization chamber at the end of 
the photon-beam line. The systematic uncertainty for the flux measurement is 
estimated below the 5\% level. The systematic uncertainty of the surface density of 
the solid $^7$Li target is estimated as 3\% (due to a somewhat irregular shape
of the target).

The largest uncertainty is related to the separation of coherent signal and breakup 
background. For coherent $\pi^0$ production we estimate a systematic uncertainty due 
to this effect of 2\% - 5\% for incident photon energies from threshold to 200~MeV,
5\% - 8\% between 200~MeV and 300~MeV, and 8\% - 20\% between 300~MeV and 500~MeV. 
For $\eta$-production most of this uncertainty is reflected in the statistical 
uncertainties of the yields, which include the uncertainty related to the fitting 
of the missing energy spectra. 

\section{Results}
\label{sec:results}

The results for the two reaction channels are of different quality and intended 
for different purposes. The $\pi^0$-data have excellent statistical quality.
In most figures their statistical error bars are smaller than the symbol sizes.
Although we compare them here only to PWIA approximations to discuss their most 
important features, they may serve as precision tests for more advanced models, 
taking into account the correct nuclear structure of $^7$Li and the nuclear effects 
beyond PWIA.

The pioneering results for $\eta$-production, at a cross section level of 10 - 20 nb, 
have limited statistical precision, but still allow a comparison of the threshold 
behavior to the $^3$He case.

\begin{figure*}[thb]
\centerline{\resizebox{0.98\textwidth}{!}{%
  \includegraphics{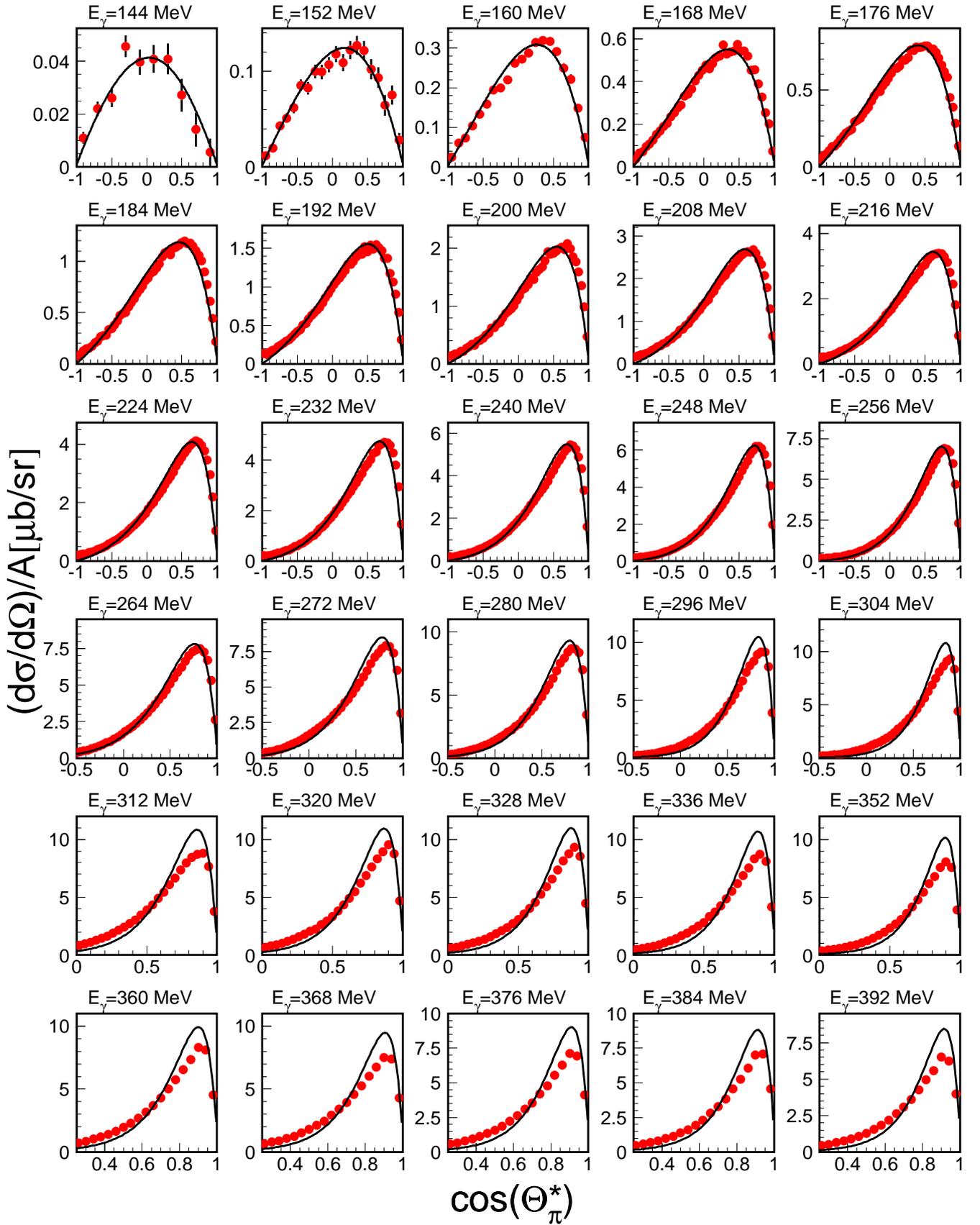}
}}
\caption{Angular distributions for coherent $\pi^0$-production for different ranges 
of incident photon energy. Curves: results of PWIA model normalized in absolute scale 
to experiment.
}
\label{fig:pi_ang}       
\end{figure*}

\clearpage

\subsection{Coherent $\pi^0$-photoproduction}
\label{sec:pires}

\begin{figure}[htb]
\resizebox{0.5\textwidth}{!}{%
  \includegraphics{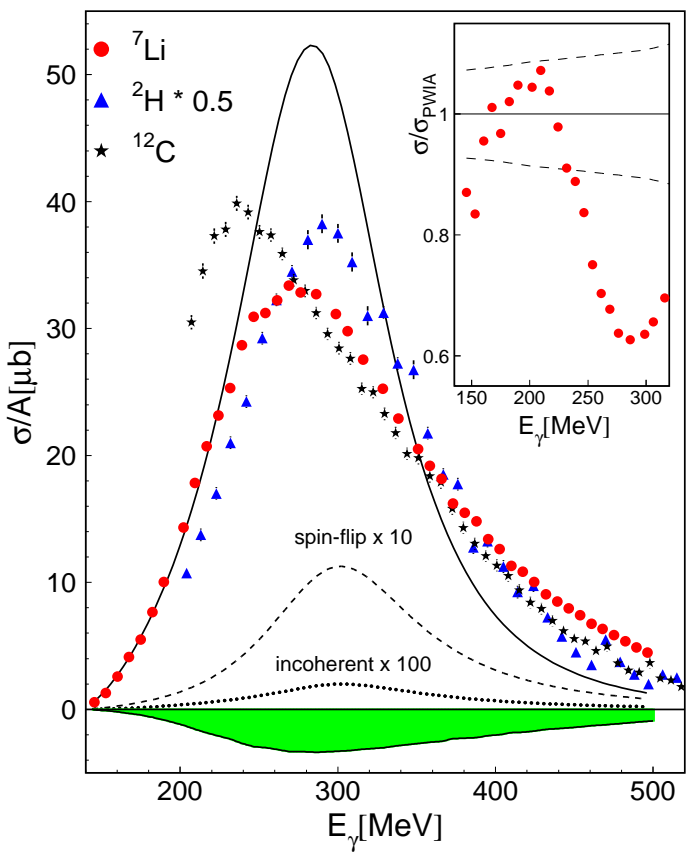}
}
\caption{Total cross section for coherent $\pi^0$-production. The shaded (green) band 
indicates the size of systematic uncertainty of the data.  
Data for the deuteron (scaled down by factor of two) \cite{Krusche_99} and $^{12}$C 
\cite{Krusche_02} for comparison.
Solid curve: PWIA results, Eq.~\ref{eq:pi_pwia}, dashed curve: predicted contribution 
of spin-flip amplitude (first term of Eq.~\ref{eq:pi_sf}) scaled up by a factor of 10, 
dotted curve: incoherent contribution from excitation of 478-keV level 
(second term of Eq.~\ref{eq:pi_sf}) scaled up by factor of 100. 
Insert: ratio of measured cross section and PWIA, dashed lines: range of systematic 
uncertainty of $^7$Li data. 
}
\label{fig:pi_total}       
\end{figure}

Angular distributions and the total cross section for the 
$\gamma+^7$Li$\rightarrow {^7}\mbox{Li}+\pi^{0}$ reaction are summarized 
in Figs.~\ref{fig:pi_ang} and \ref{fig:pi_total}. We discuss first the total cross 
section. The energy dependence and absolute magnitude reflect the properties of the 
elementary production cross section off the nucleon, trivial factors like $A^2$ and 
sin$^{2}(\Theta_{\pi}^{\star})$, the nuclear form factor, FSI effects, and possible 
in-medium modifications of the involved nucleon resonances (here the $\Delta$(1232)).
We compare the data to similar results for the deuteron \cite{Krusche_99} and 
$^{12}$C \cite{Krusche_02}. The systematic evolution of the $\Delta$-resonance peak 
in dependence of the nuclear mass number from `almost free' production for the deuteron 
to `almost nuclear density' for carbon is clearly visible. 

One should keep in mind, as discussed in detail in \cite{Krusche_02}, 
that the effective position of the $\Delta$-resonance peak is determined by different 
effects: the interplay between the nuclear form factor and the sin$^2(\Theta)$ term 
in the PWIA approximation (see Eq.~\ref{eq:pi_no}; not valid for the $J=1$ deuteron 
which is lacking the sin$^2$ term), the FSI effects in distorted-wave impulse
approximation (DWIA), and the density 
dependent in-medium modification of the position and width of the resonance. 
Actually, the model of Drechsel et al. \cite{Drechsel_99}, which reproduced quite
well the data for nuclei from carbon to lead \cite{Krusche_02}, predicts an
{\em upward} shift of the $\Delta$(1232) in-medium resonance position; although
the peak in the cross section appears to be {\em downward} shifted due to the 
other effects. The lithium case is interesting because it is transitional 
between the $\Delta$-in-vacuum and $\Delta$-in-normally-dense-matter
cases. Previous results \cite{Krusche_02} have shown that the measured cross sections
from carbon to lead can be reproduced with $\Delta$-self energies extracted from
$^{4}$He data. However, $^4$He is itself a very dense nucleus and the effective density
of $^7$Li is significantly lower than for any of the nuclei studied so far.
The extraction of $\Delta$-self energies from the lithium data will require detailed
model calculations, taking into account the FSI effects, which are not yet available
but in progress.

Here, we compare the measured cross sections to the PWIA modeling discussed in 
Sec.~\ref{sec:pwia}. It is obvious from the figure that the elastic spin-flip-term 
(term with $F^2_{C*}$ in  Eq.~\ref{eq:pi_sf}) and the incoherent excitation of 
the 478-keV state of the $^7$Li nucleus (term with $F^2_{Cx*}$ in Eq.~\ref{eq:pi_sf})
are negligible effects for the total cross section, both much smaller than 
the systematic uncertainty of the data. (Note, however, the importance of the 
spin-flip-term for the angular distributions discussed below.) 
In the low-energy range, up to incident photon energies of $\approx$ 225 MeV, the 
measured cross sections agree surprisingly well with the PWIA results (mostly within
systematic uncertainties of the data).
This demonstrates that the trivial effects of the coherent process are well understood 
in PWIA and that in this regime effects from FSI and in-medium modifications of the 
$\Delta$-resonance must be either both small or cancelling. In the maximum of the 
$\Delta$-resonance, PWIA largely overestimates the data. This is consistent with the 
expected onset of strong FSI and the in-medium damping of the $\Delta$-resonance. 
At even higher incident photon energies, beyond the energy range where the elementary 
cross section is dominated by the $\Delta$(1232) excitation, the model is missing 
contributions from other photoproduction multipoles (e.g. from the excitation of the 
P$_{11}$(1440) and D$_{13}$(1520) resonances and background terms), so that no 
agreement can be expected. 

\begin{figure}[ttb]
\resizebox{0.49\textwidth}{!}{%
  \includegraphics{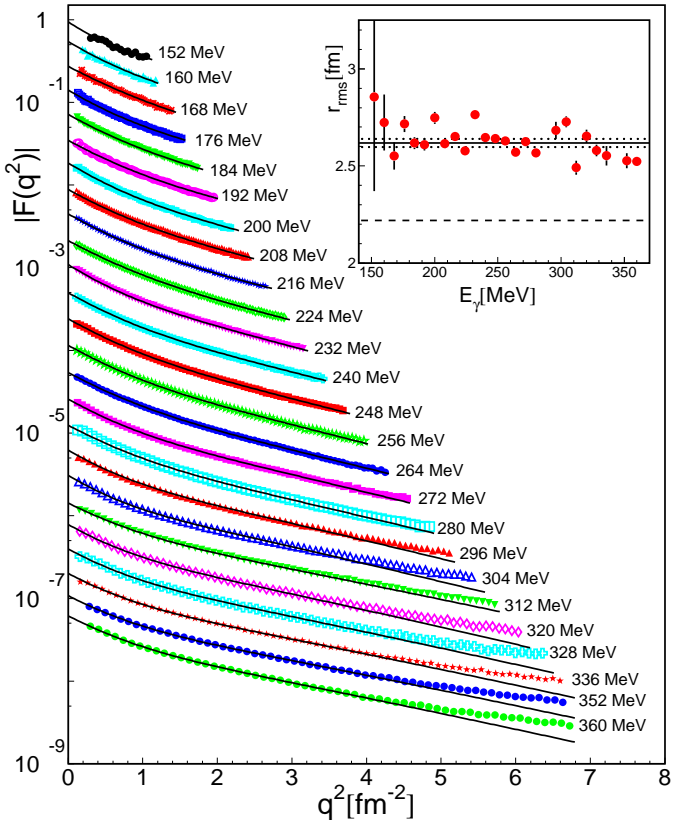}
}
\caption{Form factor of $^7$Li extracted from the ratio of measured angular 
distributions and PWIA results for different ranges of incident photon energy 
(see text). The absolute scale corresponds to the 152 MeV data, the other data are 
scaled down by successive factors of two. The solid lines correspond to fits with 
Eq.~\ref{eq:owells}. The insert shows the rms mass radii (red dots) extracted from the 
fits with Eq.~\ref{eq:mod_rms}. The solid line represents
the average (dotted lines statistical uncertainty), the dashed line the rms charge 
radius (for point-like protons). 
}
\label{fig:ff}       
\end{figure}

The shape of the angular distributions in Fig.~\ref{fig:pi_ang} is quite well reproduced 
at low incident photon energies and even reasonably well at higher energies. This is so, 
because the shape is dominated by the sin$^2(\Theta)$ term and the nuclear form factor.
However, a closer inspection of the angular distributions also shows some systematic 
deviations between experiment and PWIA for the energy range where FSI effects seem to 
be small. For a more detailed analysis Figure~\ref{fig:ff} shows a reduced version 
of the differential cross sections as a function of the squared momentum transfer 
$q^2$. The cross sections have been divided by the PWIA estimate from Eq.~\ref{eq:pi_pwia}, 
but without the form-factor terms in Eqs.~\ref{eq:pi_no} and \ref{eq:pi_sf}. The square roots 
of these ratios, shown in the figure, correspond to the nuclear mass form factor when the 
PWIA is valid (and the incoherent excitation can be neglected). Shown are only the results 
for pion cm-angles with $\mbox{cos}(\Theta^{\star}_{\pi})>-0.5$, where the PWIA 
approach seems to be reasonable. The first important observation is that the 
$q^2$-dependence of these distributions is almost independent of incident photon energy. 
This is what one would expect for a $q^2$ dependence related to the nuclear form factor. 

It was then tested whether the data can be fitted with a model of the form factor. 
The form corresponding to a simple harmonic oscillator shell model
\begin{equation}
F_{HO}(q^2) = d(1-cq^2)\mbox{exp}(-aq^2)
\label{eq:swell}
\end{equation}   
did not give satisfying results for the whole range of momentum transfers 
(see below). Much better results were obtained with the double-well form for 
$s$- and $p$-orbits used in \cite{Suelzle_67}:
\begin{eqnarray}
F_{MO} & = & a_0\left[\frac{2}{3}\mbox{exp}(-q^2b_1^2/4)\right.\nonumber\\
& & \left.+ \frac{1}{3}(1-q^2a_2^2/6)\mbox{exp}(-q^2b_2^2/4)\right]\nonumber \\
b^2_i & = & a^2_i(1-1/A),\;\;\; i=1,2
\label{eq:owells}
\end{eqnarray}
where $a_0$ accounts for the overall normalization and $a_1$, $a_2$ are the 
well-strength parameters of the $s$- and $p$-wells. Fits with this model form factor
are shown in Fig.~\ref{fig:ff} as solid lines. They excellently describe the data over 
a large range of incident photon energies and momentum transfers with almost identical
parameters. The average values of the well-strength parameters are  
\begin{eqnarray}
a_1 & = & (1.599\pm0.001)\;\; \mbox{fm}\nonumber\\
a_2 & = & (2.47\pm0.06)\;\; \mbox{fm}\;.
\end{eqnarray} 
Suelzle et al. \cite{Suelzle_67} quote for the charge distribution parameters 
$a_1$=(1.55$\pm$0.015)~fm and $a_2$=(2.02$\pm$0.06)~fm so that the s-well strength
is very similar for charge and mass distribution (3\% difference), while
the p-well strength is $\approx$~20\% larger for the mass distribution.  
The rms radius is related to these parameters by:
\begin{equation}
r^2_{rms} = \frac{A-1}{A}\left(a_1^2+\frac{1}{2}a_2^2\right)+\frac{1}{3}a_2^2 .
\label{eq:mod_rms}
\end{equation}
The values for $r_{rms}$ obtained from the fits are shown in the insert of 
Fig.~\ref{fig:ff}. They show no systematic variation with incident photon energy 
and their average of $\approx$ 2.62~fm is significantly larger than the $r_{rms}$ 
radius of the charge distribution ($\approx$ 2.27~fm in \cite{Suelzle_67}; 
note that the value of 2.43~fm quoted in this reference includes the charge radius 
of the proton).

\begin{figure}[thb]
\resizebox{0.49\textwidth}{!}{%
  \includegraphics{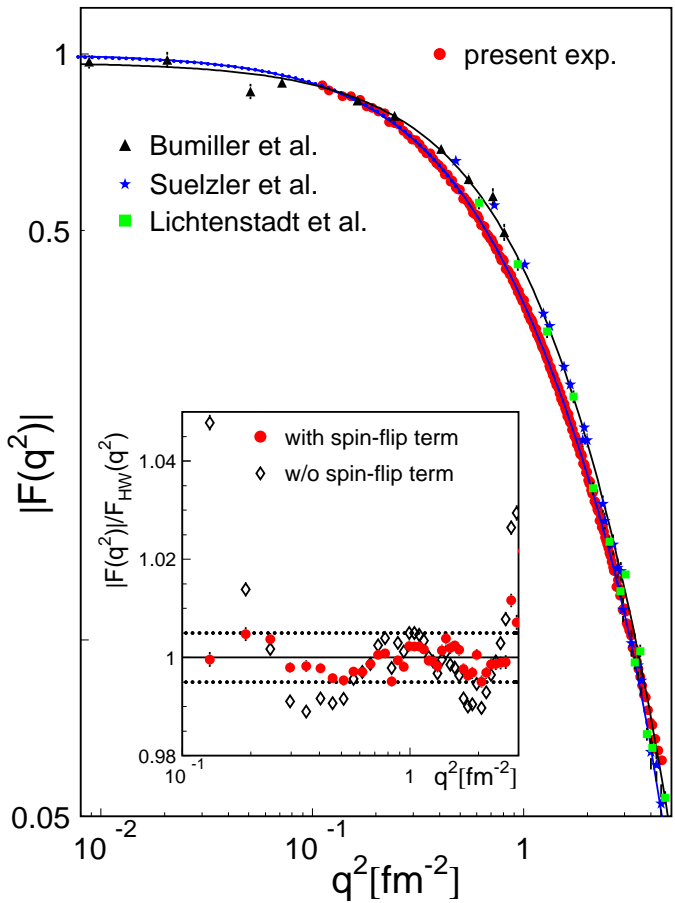}
}
\caption{Average of the form factors from Fig.~\ref{fig:ff} compared to the charge 
form factors from electron scattering (divided by proton charge form factor). 
Solid (blue) line: fit to present data 
($q^2<$ 3~fm$^{-2}$) with Eq.~\ref{eq:owells}. Results for fit with 
Eq.~\ref{eq:series} and $N=5$ ($q^2<$ 3 fm$^{-2}$), and fit with $N=3$ for 
$q^2<$0.5~fm$^{-2}$ (dashed and dotted lines) are not distinguishable from solid line; 
(black) solid line: fit of electron scattering data with Eq.~\ref{eq:owells}.
}
\label{fig:radius}       
\end{figure}

\begin{table*}[th]
  \caption[Properties of meson]{
    \label{tab_01}
Fit results for the mass rms-radius. Results are given in column (1) for fits with 
the full PWIA model (Eqs.~\ref{eq:pi_no},\ref{eq:pi_sf},\ref{eq:pi_pwia}), 
in column (2) for a truncated model without the spin-flip contribution 
(Eq.~\ref{eq:pi_sf}), and in column (3) for a model with the spin-flip contribution 
arbitrarily doubled. Column (4) shows for comparison results of fits to charge form 
factor from electron scattering (divided by proton charge form factor). 
First row $^{1)}$ average of the fit results from Fig.~\ref{fig:ff} with the 
double-well model (Eq.~\ref{eq:owells}) over an range of incident photon energies 
from 150 MeV - 360 MeV ($\chi^2$ values are averages for all fits). Second row $^{2)}$ 
fit with the double-well model to the 
averaged form factor for $q^2<$ 3 fm$^{-2}$ (Fig.~\ref{fig:radius}). Third row $^{3)}$ 
fit with series (Eq.~\ref{eq:series} with $N=5$ (for the model with neglected spin-flip
term with $N=7$, since $N=5$ did not converge). Fourth row $^{4)}$ fit with series 
with $N=2$ for $q^2<$ 0.5 fm$^{-2}$ (only few data for charge form factor).
}
  \begin{center}
    \begin{tabular}{|c|c|c|c|c|c|c|c|c|}
      \hline 
      & \multicolumn{2}{c|}{full model}
      & \multicolumn{2}{c|}{no spin-flip}
      & \multicolumn{2}{c|}{spin-flip doubled}
      & \multicolumn{2}{c|}{charge form factor}\\ 
      method       
      & $r_{rms}^{(m)}$ [fm] 
      & $\chi^2$ 
      & $r_{rms}^{(m)}$ [fm] 
      & $\chi^2$ 
      & $r_{rms}^{(m)}$ [fm] 
      & $\chi^2$ 
      & $r_{rms}^{(ch)}$ [fm] 
      & $\chi^2$\\            
      \hline  
    $^{1)}$  double well     & 2.618$\pm$0.004 & 3.4 & 2.710$\pm$0.004 & 7.2 & 2.587$\pm$0.004 & 3.2 & - & -\\
    $^{2)}$  double well     & 2.659$\pm$0.007 & 8.6 & 2.898$\pm$ 0.003 & 24 & 2.612$\pm$0.002 & 8.4 & 2.30$\pm$0.02 & 3.7\\
    $^{3)}$  series N=5 (7)  & 2.635$\pm$0.002 & 8.4 & 2.981$\pm$ 0.002 & 16  & 2.575$\pm$0.002 & 8.3 & 2.17$\pm$0.04 & 1.9\\ 
    $^{4)}$  series N=2 ($q^2<$ 0.5 fm$^{-2}$) & 2.56$\pm$0.12 & 5.9 & 3.12$\pm$0.08 & 14 & 2.398$\pm$0.15 & 7.2 & 2.2$\pm$1.2 & 1.2\\
      \hline
    \end{tabular}
  \end{center}
\end{table*}

For a more detailed analysis the average of the distributions from Fig.~\ref{fig:ff}
for incident photon energies up to 280 MeV (after renormalization of their absolute 
scales) is compared in Fig.~\ref{fig:radius} to the charge form-factor values from 
\cite{Suelzle_67,Bumiller_72,Lichtenstadt_89}. It is evident that the 
$q^2$-de\-pen\-dence of the electron scattering data is different from the present 
results. Both data sets have been fitted for the range of $q^2< 3$~fm$^{-2}$
with Eqs.~\ref{eq:swell} and \ref{eq:owells}. The fits with the simple harmonic 
oscillator model (Eq.~\ref{eq:swell}) were of much inferior quality 
(reduced $\chi^2 \approx$ 880 for present data compared to $\approx$ 8 for the 
double-well form Eq.~\ref{eq:owells}) and were not further considered. The fits with 
the double-well form from Eq.~\ref{eq:owells} are shown in Fig.~\ref{fig:radius} 
as solid blue (present data) and solid black (electron scattering data) lines. 
They correspond to the following rms-radii ($r_{rms}^{(ch)}$: electron data, 
$r_{rms}^{(m)}$: present data):
\begin{eqnarray}
r_{rms}^{(ch)} & = & (2.30 \pm 0.02)\;\; \mbox{fm},\\
r_{rms}^{(m)}    & = & (2.66 \pm 0.01)\;\; \mbox{fm}.
\end{eqnarray}

The insert of Fig.~\ref{fig:radius} shows the ratio of the present data and this 
fit (filled, red points). For $q^2$-values up to 3~fm$^{-2}$ the fit reproduces the 
shape to within $\pm$0.5\%. Due to this small systematic differences between fit
curve and data, the result for the radius is almost independent on the fitted range.
If, for example, we fit only the data for incident photon energies below 225 MeV,
where agreement between data and PWIA is best, the radius changes only from 2.659~fm
to 2.653~fm.  
Also shown in the insert (black, open points) is the 
result from an analysis that neglected the spin-flip term (Eq.~\ref{eq:pi_sf}) in 
the elementary production cross section. The influence of this term is substantial 
at small $q^2$ values; the reduced $\chi^2$ of the fit rises from 8.6 to 24 if it is
omitted. 

The rms radius can be also extracted from the present data without the use of a 
specific model for the form factor from its slope for $q^2\rightarrow 0$, using 
the expansion
\begin{equation}
F(q^2) = 1 - \frac{q^2}{6} r^2_{rms} + {\cal{O}}(q^4).
\end{equation}
The data were fitted with the ansatz
\begin{equation}
F(q^2) = \sum_{n=0}^{N}c_nq^{2n}\;,
\label{eq:series}
\end{equation}
from which the rms radius follows as
\begin{equation}
r_{rms} = \sqrt{-6c_1/c_0}\;,
\label{eq:rms}
\end{equation}
where for correctly normalized form factors $c_0$ would be unity (here it differs 
by a few per cent from unity).

Different fits have been exploited. Two extreme cases are fits for the $q^2$ range
up to 3~fm$^{-2}$ with $N=5$ and with $N=2$ only for small momentum transfers 
($q^2 <$ 0.5~fm$^{-2}$). The results for $r_{rms}^{(m)}$ extracted from Eq.~\ref{eq:rms} 
are in agreement and close to the above value from the double-well harmonic-oscillator 
model: 
\begin{equation}
r_{rms}^{(m)}= (2.635\pm 0.002)\;\;\mbox{fm} 
\end{equation}
for the $N=5$ fit over the full range and
\begin{equation}
r_{rms}^{(m)} = (2.56 \pm 0.12)\;\; \mbox{fm}
\end{equation}
for the slope from the low-momentum transfer $N=2$ fit (quoted uncertainties are 
statistical). The fit curves are so similar to the double-well result
that they are indistinguishable from it in Fig.~\ref{fig:radius}.

The results for all model fits are summarized in Table~\ref{tab_01}. The form factors
derived in PWIA from the coherent pion data correspond to an $rms$-mass radius of
$\approx$ (2.60 - 2.65) fm$^{-2}$ (column (1) of the table), which is significantly 
larger than the result for the charge radius (column (4) of the table).
The reduced $\chi^2$ of the fits is larger than unity, which is due to the
systematic structure of the form factor at the sub-percent level (see insert of
Fig.~\ref{fig:radius}), which is significant within statistical uncertainties,
but much smaller than (energy dependent) systematic uncertainties not included 
in the fitting process.

One source of systematic uncertainty is the contribution of
the spin-flip-term (Eq.~\ref{eq:pi_sf}) in the PWIA approximation, which includes 
only the dominant $M_{1+}$ amplitude and ignores all other production multipoles. 
Its influence has been tested by model fits: excluding it completely (column (2) of 
Table~\ref{tab_01}) and arbitrary doubling its strength (column (3)). 
Excluding the spin-flip contribution increases significantly the $\chi^2$ of the fits 
and increases the value of the radius. Enhancing the spin-flip term by a factor of 
two is quite a large (probably unrealistic) variation, since in the $\Delta$-resonance 
range it is strongly dominated by the well-known $M_{1+}$ multipole. 
The $\chi^2$-values of these fits are similar to the standard version and the radius 
becomes smaller, but is still larger than the charge radius. The comparison gives 
some indication of the possible size of systematic uncertainty due to this term.   

The contribution of the inelastic 478-keV excitation was ignored for the form-factor 
extraction. To test its importance we subtracted the PWIA estimate for this 
process from the measured angular distributions and repeated the analysis.
This removes strength at large $q^2$, which makes the form factor steeper and thus 
tends to increase the radius.  However, the effect is smaller than statistical 
uncertainties and can be safely neglected.

\begin{figure}[thb]
\resizebox{0.49\textwidth}{!}{%
  \includegraphics{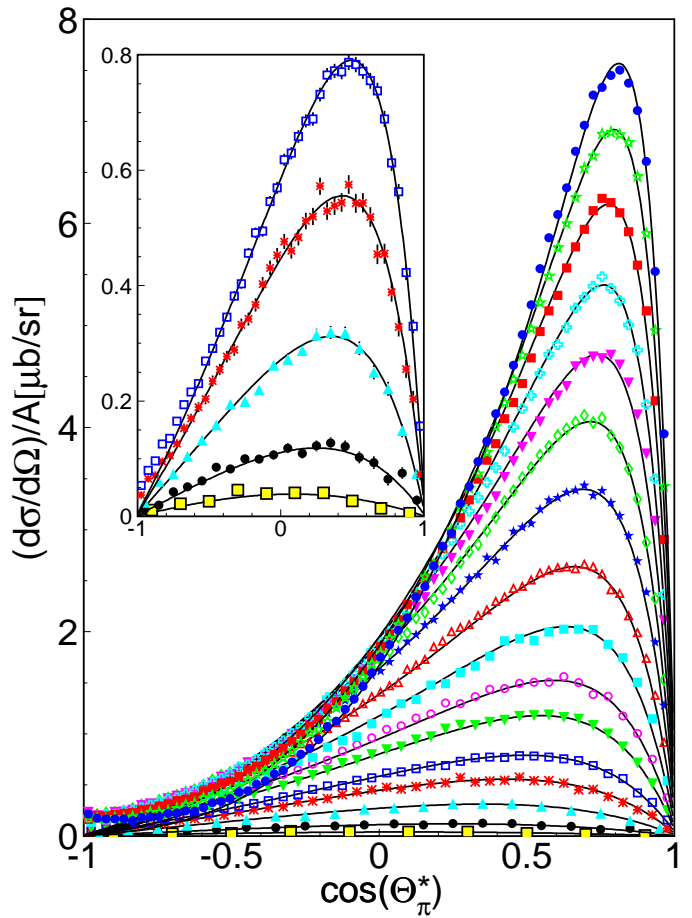}
}
\caption{ Main picture: angular distributions for incident photon energies from 
152 MeV (bottom curve) to 264 MeV (top curve, same energy bins as in 
Fig.~\ref{fig:ff}) compared to PWIA results using the 
fitted form factor. Absolute scales of model results normalized to data (see text). 
The insert shows on a larger scale the low-energy results (144 MeV - 168 MeV).
}
\label{fig:difflin}       
\end{figure}

So far no model results are available for FSI effects in $^7$Li. 
Results for other light nuclei ($^4$He, $^{12}$C) \cite{Drechsel_99,Krusche_02}
have shown that they are important for the energy dependence of the
total cross section.  
Nevertheless, the good agreement of the measured total cross section with the PWIA
modelling at incident photon energies below 225 MeV indicates that they must
be small for $^7$Li in this energy range. The main FSI effect depends on the pion 
kinetic cm energies (and thus on the incident photon energy) but it could also modify 
to some extent the shape of the angular distributions, which are the basis for the
form-factor extraction. However, the form-factor fits (see insert of Fig.~\ref{fig:ff})
give consistent results for the mass radius $r_{rms}^{(m)}$ over a wide range of incident
photon energy, over which the energy-de\-pen\-dent FSI effects change drastically,
from a few per cent between 180 MeV and 220 MeV to almost 40\% around 280 MeV
(see insert of Fig.~\ref{fig:pi_total}).

In order to explain the observed difference between the extracted form factor and the 
charge form-factor data, FSI effects with a very peculiar behavior would be needed.
This is demonstrated in Fig.~\ref{fig:difflin}, where the low-energy angular 
distributions are compared to a modified PWIA. The only difference to the PWIA
curves in Fig.~\ref{fig:pi_ang} is that instead of the form factor from electron
scattering the double-well parameterization of the present form factor data
from Fig.~\ref{fig:radius}, corresponding to $r_{rms}^{(m)}$=2.66~fm was used.
The absolute scales of the PWIA results were renormalized to the data in order to
remove the energy-de\-pen\-dent FSI effects. This PWIA must describe by construction 
the angular distributions on average. However, it actually agrees almost 
perfectly with the shapes of all individual distributions, with very different
relations between pion angles and nuclear momentum transfers. This means that an FSI 
effect would be needed, which over a range of incident photon energy of more than 
100 MeV has exactly the same angular and momentum-transfer dependence as a change 
of the form factor from an rms radius of 2.3 fm to 2.66 fm. Although this does not
seem to be a likely scenario, reasonably sophisticated modelling of the FSI 
effects is needed before a final conclusion can be drawn. However, results for
a similar analysis of coherent photoproduction off carbon, calcium, and lead nuclei
point to a small influence of FSI on the extracted radii for light nuclei. 
Fully taking into account the FSI effects \cite{Krusche_05} lowered the extracted
value of the mass radius for $^{208}$Pb by 5.8\%, for $^{40}$Ca by 2.2\%, 
but for the lighter $^{12}$C only by 0.9\%, while the observed difference between
charge and mass radius for $^7$Li is on the 10\% level.

\subsection{Coherent $\eta$-photoproduction}

\begin{figure}[thb]
\resizebox{0.49\textwidth}{!}{%
  \includegraphics{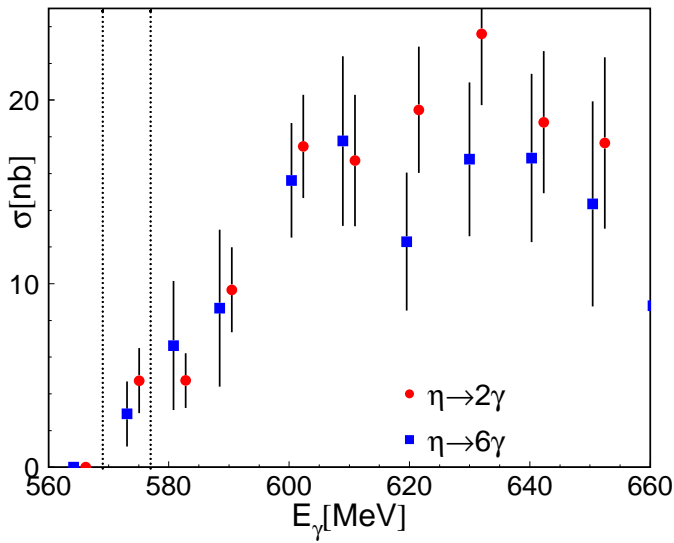}
}
\caption{Comparison of the total cross section for coherent $\eta$-production from 
the two-photon and six-photon decay of the $\eta$. The vertical dotted lines 
indicate coherent and breakup thresholds. 
}
\label{fig:eta_total}       
\end{figure}

\begin{figure}[htb]
\resizebox{0.49\textwidth}{!}{%
  \includegraphics{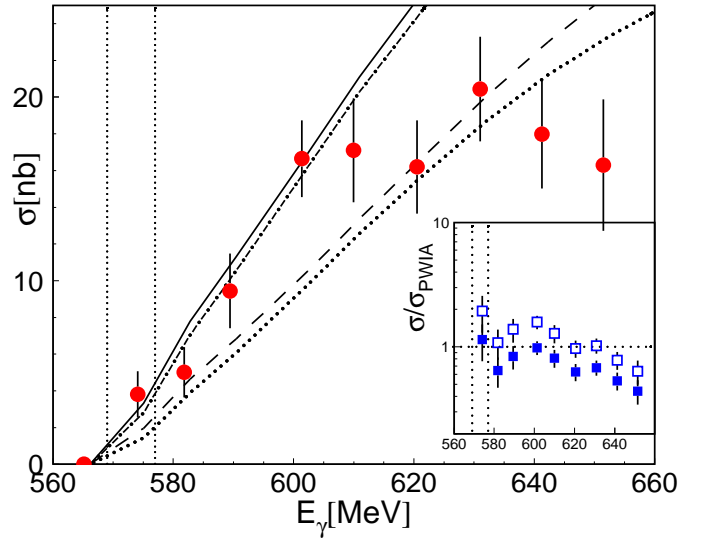}
}
\caption{Comparison of the average of the experimental two-photon and six-photon 
cross sections to the PWIA results. Dashed (dotted) curves: coherent contribution
(Eq.~\ref{eq:eta} without $F^2_{Cx*}$-term) based on charge form factor 
(mass form factor fitted to pion production).
Solid (dash-dotted) curves: sum of coherent and incoherent contribution (see Eq.~\ref{eq:eta})
for charge (mass form factor).
The insert shows the ratio of measured cross section and PWIA results,
open symbols only coherent part, filled symbols sum of coherent and incoherent
contributions).  
}
\label{fig:eta_pwia}       
\end{figure}

The total cross sections extracted for the two $\eta$-decay channels, summarized 
in Fig.~\ref{fig:eta_total}, are nicely consistent. They show a much smoother rise 
at production threshold than the $^3$He data (cf. Fig.~\ref{fig:helium}).
For a quantitative analysis their average is compared to PWIA modelling,
based on Eq.~\ref{eq:eta} in Fig.~\ref{fig:eta_pwia}. As discussed in 
Sec.~\ref{sec:pwia} the situation is much different to pion production since for 
$\eta$-production the cross section is dominated by the contribution of the odd 
$1p_{3/2}$ proton, which is only a small correction in the $\pi^0$ case.

\begin{figure}[ttb]
\resizebox{0.49\textwidth}{!}{%
  \includegraphics{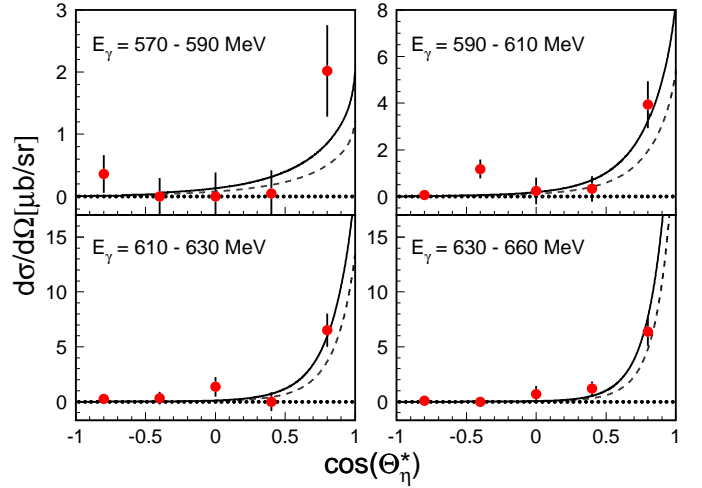}
}
\caption{Comparison of measured angular distributions for 
$\gamma$+$^7$Li$\rightarrow ^7$Li+$\eta$
{(red) dots} to PWIA results, solid lines: full PWIA, dashed lines: only coherent
part. 
}
\label{fig:eta_diff}       
\end{figure}

Results from PWIA, using the charge form factor or the mass form factor fitted 
to the pion data (dashed, respectively dotted curves in Fig.~\ref{fig:eta_pwia}),
are similar. This is simply so because, for $\eta$ production, large momentum 
transfers dominate where the two form factors agree. The relative contribution
of the incoherent excitation of the 478-keV state is significant in PWIA.
Also this had to be expected because for $\eta$-production there is no piece 
with an $A^2$-term like Eq.~\ref{eq:pi_no} for coherent pion production and the 
elastic and inelastic form factors are similar for large momentum transfers. 
The systematic uncertainty of the PWIA results is larger than in
the pion case because the cross section is dominated by these less well-established
contributions. However, altogether the comparison of the energy dependence of the
measured total cross section and the PWIA results in Fig.~\ref{fig:eta_pwia}
shows no threshold enhancement above phase-space behavior, and thus no
indication for the formation of a quasi-bound state. 
The situation is thus much different from the $^3$He case discussed in the
introduction which, apart from the incoherent excitation, has similar systematic
uncertainties in PWIA. Comparison of the two results highlights the special role 
of the $\eta-^3\mbox{He}$ system. 

Also the results for the angular distributions, summarized in Fig.~\ref{fig:eta_diff},
are consistent with this interpretation. They agree better with the momentum-transfer
dependence of the form factor than in the $^3$He case and show no tendency towards
isotropic behavior close to threshold.

\section{Summary and Conclusions}

Precise data have been measured for coherent photoproduction of $\pi^0$-mesons 
off $^7$Li nuclei and coherent photoproduction of $\eta$-mesons off the same nucleus 
has been identified for the first time. The experimental results for the pion 
production are quite well reproduced at low incident photon energies by a PWIA 
dominated by the spin/isospin-independent part of the elementary production 
amplitude. The spin-flip amplitude from the unpaired $1p_{3/2}$ proton is considered 
for the leading $M_{1+}$ multipole and the corrections applied for the incoherent 
excitation of the 478-keV nuclear state in $^7$Li are insignificant. This model 
reproduces quite well total cross sections and angular distributions at incident 
photon energies below 225 MeV, indicating that distortion effects from final-state 
interactions of the pion are small in this energy range.

After an adjustment of the nuclear form factor, which corresponds to a change
in the harmonic double-well parameterization from an rms radius of 2.3 fm 
(reported for the charge form factor derived from electron scattering data) to 
2.66 fm, the shape of angular distributions in this energy range is excellently
reproduced. 
Exploiting the possible uncertainties due to approximations, in particular in the
spin-flip term of the PWIA, we find reasonable agreement between data and PWIA
for rms radii down to 2.5 fm, which are still larger than previously reported charge 
radii and also larger than predictions for the mass radius, which are around 2.35 fm 
(see e.g. \cite{Tomaselli_00,Kajino_88}). DWIA calculations with careful treatment 
of possible FSI effects are needed for further analysis of this discrepancy.  

Coherent photoproduction of $\eta$-mesons is quite difficult to measure and so far 
only results for the deuteron \cite{Hoffmann_97,Weiss_01} and $^3$He 
\cite{Pfeiffer_04,Pheron_12} had been reported. This experiment extended the mass 
range to $^7$Li by measuring total cross sections on the level below 20 nb. 
The results, also for the angular distributions, are in good agreement with 
PWIA expectations and do not show an unexplained threshold enhancement as in the 
$^3$He case, underlining the special role of $^3$He as a candidate for $\eta$-mesic 
states.

\vspace*{1cm}
{\bf Acknowledgments}

We wish to acknowledge the outstanding support of the accelerator group 
and operators of MAMI. We thank L. Tiator for the discussion of the plane wave approximations.
This work was supported by Schweizerischer Nationalfonds, Deutsche
Forschungsgemeinschaft (SFB 443, SFB/TR 16), DFG-RFBR (Grant No. 05-02-04014),
UK Science and Technology Facilities Council, STFC, European Community-Research 
Infrastructure Activity (FP6), the US DOE, US NSF and NSERC (Canada).

\end{document}